\newif\ifpp
\ppfalse
\ifpp
\documentclass[11pt,a4paper]{article}
\else
\documentclass{svjour3}                     % onecolumn (standard format)
\usepackage[figuresleft]{rotating}

\fi
\pdfoutput=1
\usepackage{graphicx}
\usepackage{todonotes}
\usepackage{siunitx}
\usepackage{tabularx,booktabs} 
\usepackage{authblk}
\usepackage{hyperref}

\newcommand{\dd}{\mathrm{d}}

\ifpp
\fi
\newcommand{\avrg}[1]{\left<  #1 \right>}

\title{Statistical sensitivity estimates for oscillating electric dipole moment measurements in storage rings
\footnote{accpetaed for publication in European Physical Journal C}}

\ifpp
\else
\author{
J\"org Pretz$^{a,b,c}$\thanks{Corresponding author.} and Fabian M\"uller$^{a,b}$\\
\llap{$^a$}Institut f\"ur Kernphysik, Forschungszentrum J\"ulich, 52425 J\"ulich, Germany\\
\llap{$^b$}III. Physikalisches Institut B, RWTH Aachen University, 52056 Aachen, Germany\\
\llap{$^c$}JARA-FAME (Forces and Matter Experiments), Forschungszentrum
J\"ulich und RWTH Aachen University\\
  E-mail: \email{pretz@physik.rwth-aachen.de}
}

\fi

\begin{document}

%\begin{center}
%{\it 
%\end{center}

\ifpp
\else

\author{J. Pretz \and S.P.~Chang \and V.~Hejny \and S. Karanth \and S. Park \and Y.~Semertzidis
\and E.~Stephenson  \and H.~Str\"oher}
\institute{J\"org Pretz \at
            Institut f\"ur Kernphysik, Forschungszentrum J\"ulich, 52425 J\"ulich, Germany \\
        III. Physikalisches Institut B, RWTH Aachen University, 52056 Aachen, Germany \\
            JARA-FAME, Forschungszentrum J\"ulich und RWTH Aachen University \\
         \email{pretz@physik.rwth-aachen.de}  \and
           S.P.~Chang  \at
           Center for Axion and Precision Physics Research, IBS, Daejeon 34051, Republic of Korea \\
           Department of Physics, KAIST, Daejeon 34141, Republic of Korea  \and
           V.~Hejny \at
           Institut f\"ur Kernphysik, Forschungszentrum J\"ulich, 52425 J\"ulich, Germany \and
           S. Karanth \at
           Institute of Physics, Jagiellionian Univsersity, Cracow, Poland \and
           S. Park  \at
           Center for Axion and Precision Physics Research, IBS, Daejeon 34051, Republic of Korea \and
            Y.~Semertzidis \at
           Center for Axion and Precision Physics Research, IBS, Daejeon 34051, Republic of Korea, \\
           Department of Physics, KAIST, Daejeon 34141, Republic of Korea \and
           E. Stephenson \at
           Indiana Univ., Bloomington, IN 47408, USA  \and
           H.~Str\"oher \at
           Institut f\"ur Kernphysik, Forschungszentrum J\"ulich, 52425 J\"ulich, Germany \\
           JARA--FAME (Forces and Matter Experiments), Forschungszentrum J\"ulich and RWTH Aachen University, Germany 
}
\fi

\date{}

\authorrunning{J. Pretz et al.}
\titlerunning{Statistical sensitivity estimates \dots}
\maketitle

\begin{abstract}
In this paper analytical expressions are derived to describe the spin motion of a particle in magnetic and electric fields
in the presence of an axion field causing an oscillating electric dipole moment (EDM).
These equations are used to estimate statistical sensitivities for axion searches at storage rings.

The estimates obtained from the analytic expressions are compared to numerical estimates from simulations
in reference~\cite{Chang:2019poy}. A good agreement is found.
\ifpp
\bigskip \noindent
%{\bf keywords:}
%data analysis, axion, spin dynamics
\else

%\PACS{07.05.Kf}
\fi

\end{abstract}

\section{Introduction and motivation}

%******************************

%Dark matter is ...
%Different observables have been proposed {graham}
%Among these ...

Axions and axion like particles (ALPs) are candidates for dark matter.
There is thus a huge experimental effort for the search of these
kind of particles. For a detailed review, we refer the reader to references~\cite{Graham:2015ouw,Irastorza:2018dyq}.
Axions and ALPs can interact with ordinary matter in various ways.
Reference~\cite{Graham:2013gfa} identifies three terms:
\begin{equation}\label{eq:coupling}
\frac{a}{f_0} \, F_{\mu \nu} \tilde{F}_{\mu \nu}, \quad
\frac{a}{f_a} \, G_{\mu \nu} \tilde{G}_{\mu \nu}, \quad
\frac{\partial_\mu a}{f_a} \bar{\Psi}_f \gamma^\mu \gamma_5  \Psi
\end{equation}
describing the coupling to photons, gluons and to the spin of fermions,
respectively.
The vast majority of experiments makes use of the first term
(e.g. Cavity experiments (ADMX), helioscopes (CAST), light-through-wall
experiments (ALPS)). In addition, astrophysical observations  also provide sensitive limits to the axion-photon coupling.
In general, it is rather difficult for these experiments
to reach masses below $10^{-6}\, \si{eV}$, one reason being that
the axion wave length becomes too large.
Furthermore, these experiments are measuring rates proportional 
to at least a small amplitude squared.

For the second (and third) term in the list~(\ref{eq:coupling}) this is different.
It turns out that the second term has the same structure as the QCD-$\theta$ term which is also responsible 
for an electric dipole moment (EDM) of nucleons. 
The axion field gives rise to an effective time-dependent $\theta$-term and oscillates at a frequency proportional to the mass of the axion $m_a$. This gives rise to an oscillating EDM. New opportunities to search for axions/ALPs with much higher sensitivity arise, because
the signal is proportional to an amplitude $A$ and not to its square. To date, NMR based methods are being used 
to look at oscillating EDMs~\cite{Budker:2013hfa}. 

%Hans:
%Storage ring EDM experiments thus provide another opportunity to %search for axions/ALPs in addition to its possibility to investigate %static electric dipole moments of charged particles [6].

Another possibility is to search for axions/ALPs in storage rings.
Storage ring experiments have been proposed to search for electric dipole moments of charge particles~\cite{Anastassopoulos:2015ura,Abusaif:2019gry}.
These experiments
allow also, with small modifications, to search for oscillating EDMs.
This possibility is  discussed in this paper. Section~\ref{sec:bmt} 
explains the principle of the experiment, how the (oscillating) 
EDM alters the spin motion in electromagnetic fields and leads 
to a polarization observable.
In section~\ref{sec:staterr} statistical sensitivities for oscillating EDMs based on these polarization are presented.

%...EDM leads to Build-up

\section{Spin motion in storage rings}\label{sec:bmt}
The spin motion relative to the momentum vector in electric and magnetic fields is governed
by the Thomas-BMT equation
~\cite{Bargmann:1959gz,Nelson:1959zz,Fukuyama:2013ioa}:
\begin{eqnarray}
\frac{d \vec{S}}{dt} &=& (\vec{\Omega}_{\mathrm{MDM}} + \vec{\Omega}_{\mathrm{EDM}}) \times \vec{S},  \label{eq:tbmt}\\  
\vec{\Omega}_{\mathrm{MDM}} &=& -\frac{q}{m} ~ \left[G \vec{B} 
%-\frac{\gamma G}{\gamma+1} \vec{\beta} \left(\vec{\beta} \cdot \vec{B} \right)
 - \left(G-\frac{1}{\gamma^2-1} \right) \frac{\vec{\beta} \times \vec{E}}{c}\right], \label{eq:ommdm}\\
\vec{\Omega}_{\mathrm{EDM}} &=& -\frac{\eta q}{2 m c} \left[\vec{E}
% - \frac{\gamma}{\gamma+1} \vec{\beta} \left(\vec{\beta} \cdot \vec{E}\right)
+ c \vec{\beta} \times \vec{B} \right]. \label{eq:omedm}
\end{eqnarray}
$\vec{S}$ in this equation denotes the spin vector in the particle rest frame, $t$ the time
in the laboratory system, $\beta$ and $\gamma$ the relativistic Lorentz factors, and $\vec{B}$ 
and $\vec{E}$ the magnetic and electric fields in the laboratory system, respectively.
The magnetic dipole moment $\vec \mu$ and electric dipole moment $\vec d$
both pointing in the direction of the particle's spin $\vec{S}$
are related to the dimensionless
quantities $G$ (magnetic anomaly) and $\eta$ in equation~\ref{eq:tbmt}:
\begin{equation}
\vec{\mu} = g \frac{q \hbar}{2 m} \vec{S} = (1+G) \frac{q \hbar}{m} \vec{S}\,
\quad \mbox{and} \quad \vec{d} = \eta \frac{q \hbar}{2 m c} \vec{S} \, .
\label{eq1}
\end{equation}

\begin{figure}
\includegraphics[width=0.7\textwidth]{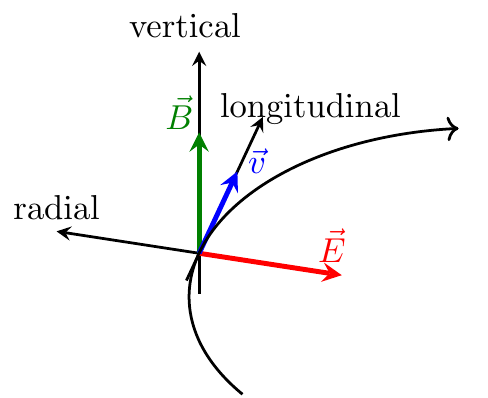}
\caption{Illustration of the coordinate system used.\label{fig:coord_sys}}
\end{figure}

We assume a vertical magnetic field and a radial electric field, constant in time, forcing the particle on a circular orbit.
The three vectors $\vec B$, $\vec E$ 
and $\vec v = \vec \beta c$ are thus mutually  orthogonal, as indicated in figure~\ref{fig:coord_sys}.
In this case
\begin{equation}
\vec \Omega_{\mathrm{MDM}} =
\left(
\begin{array}{c}
0\\
\Omega_{\mathrm{MDM}} \\
0\\
\end{array}
\right) \quad \mbox{and} \quad
\vec \Omega_{\mathrm{EDM}} = 
\left(
\begin{array}{c}
\eta \tilde \Omega_{\mathrm{EDM}} \\
0 \\
0 \\
\end{array}
\right) 
\end{equation}
with 
$\Omega_{\mathrm{MDM}} = - \frac{q}{m} (G B +  \left(G-\frac{1}{\gamma^2-1} \right) \frac{\beta E}{c})$
and
$\tilde \Omega_{\mathrm{EDM}} = -\frac{q}{2mc} (E + c \beta B)$,
 $B=|\vec B|$ and $E=|\vec E|$.
The coordinate system is chosen such that the first component points in radial direction, 
the second in vertical
and the third in longitudinal direction.
Note that $\vec \beta \times \vec E$ is anti-parallel to $\vec B$. This explains the $ +  $ sign
in front of $ \left(G-\frac{1}{\gamma^2-1} \right)$ in the definition of $\Omega_{\mathrm{MDM}}$
instead of a $- $ sign in equation~\ref{eq:ommdm}.

For the following discussion it is more convenient
to write equation~\ref{eq:tbmt} in matrix form:
\begin{equation}\label{eq:dgl}
\frac{d \vec{S}}{dt} = (A_{\mathrm{MDM}} + A_{\mathrm{EDM}})  \vec{S}
\end{equation}
with (to simplify the notation we use $\Omega_{\mathrm{EDM}}$ instead of  $\tilde \Omega_{\mathrm{EDM}}$ from now on)
\begin{equation}\label{eq:dgl1}
A_{\mathrm{MDM}} = 
\left(
  \begin{array}{ccc}
0 & 0 & \Omega_{\mathrm{MDM}} \\  
0 & 0 & 0 \\
-\Omega_{\mathrm{MDM}} & 0 & 0\\
\end{array}
  \right)
\quad \mbox{and} \quad 
A_{\mathrm{EDM}} = \eta
\left(
  \begin{array}{ccc}
0 & 0 & 0 \\  
0 & 0 & \Omega_{\mathrm{EDM}} \\
0 & -\Omega_{\mathrm{EDM}} & 0 \\
\end{array}
  \right) \, .
\end{equation}

In the following we assume that the EDM can have a constant term and a time varying component,
thus $\eta = \eta_0 + \eta_1 \cos(\omega_a t + \varphi_a)$ as suggested in ~\cite{Graham:2013gfa,Graham:2011qk}.
The oscillating term is caused by an axion of mass given by the relation $\omega_a = m_a c^2/\hbar$.
Assuming $\eta_0, \eta_1 \ll G$, $A_{\mathrm{EDM}}$ in equation~\ref{eq:dgl} can be treated as an perturbation.

The solution to first order in $\eta_0$ and $\eta_1$ for arbitrary initial condition of the spin is given in Appendix~\ref{app:soldgl}.
The best sensitivity to $\eta_0$ and $\eta_1$ is obtained by observing
a build-up of a vertical polarization of a beam  initially polarized 
in the horizontal plane. 
Thus we are interested in the behavior of the vertical spin component $S_v(t)$ 
in the case where the spin points for example initially in the longitudinal direction ($\vec S(0) = (0,0,1)^T$). 
Using equation~\ref{eq:A23} in Appendix~\ref{app:soldgl} one finds:  
\begin{eqnarray}
  S_v(t) &=&  \eta_0  \Omega_{\mathrm{EDM}}   \frac{\sin(\Omega_{\mathrm{MDM}} t)}{ \Omega_{\mathrm{MDM}}} 
    + \eta_1 \, \frac{\Omega_{\mathrm{EDM}}}{2 (\omega_a - \Omega_{\mathrm{MDM}} ) (\Omega_{\mathrm{MDM}} + \omega_a)} 
    \nonumber \\
    &&\Big[ -2 \omega_a \sin(\varphi_a)  \nonumber \\
&&  +  (\omega_a + \Omega_{\mathrm{MDM}})
  \sin( (\omega_a - \Omega_{\mathrm{MDM}}) t +\varphi_a ) 
    \nonumber \\
 &&    +  (\omega_{a} - \Omega_{\mathrm{MDM}})
  \sin((\Omega_{\mathrm{MDM}} + \omega_a) t+ \varphi_a) \Big] \, . \label{eq:sy}
\end{eqnarray}

%For the resonance case where the EDM frequency equals the spin revolution frequency $\omega_a =  \Omega_{\mathrm{MDM}}$, this reduces to
%\begin{equation}\label{eq:sy1}
%S_v(t) =  \frac{\Omega_{\mathrm{EDM}}}{2\omega_a}  \left(2 \eta_0  \sin(\omega_a t) + \eta_1 \left( \omega_a  t \cos(\varphi_a) + 
 %  \cos(\omega_a t+\varphi_a) \sin(\omega_a t) \right) \right) \, .
%\end{equation}

We are interested in the behavior close to the resonance condition
$\Omega_{\mathrm{MDM}} \approx \omega_a$.
Ignoring in equation~\ref{eq:sy} all fast oscillating terms
 (i.e. assuming $\Omega_{\mathrm{MDM}},(\Omega_{\mathrm{MDM}}+\omega_a) \gg  \Omega_{\mathrm{MDM}}-\omega_a)$ one finds
\begin{eqnarray}
S_v(t) &=&   
 \frac{\eta_1 \Omega_{\mathrm{EDM}}}{2 (\omega_a -\Omega_{\mathrm{MDM}} ) }  \, \Bigg( - { \sin(\varphi_a) } 
+  
\sin\left( (\omega_a -\Omega_{\mathrm{MDM}}) t +\varphi_a \right) \Bigg) \, .
     \label{eq:sy2}\\
     &=& \eta_1 \frac{\Omega_{\mathrm{EDM}}}{2\Delta \omega}
     \left( -\sin(\varphi_a) + \sin(\Delta \omega t + \varphi_a)\right)  \label{eq:sy2a}
\end{eqnarray}
with $\Delta \omega = \omega_a - \Omega_{\mathrm{MDM}}$
For $\varphi_a=0$ this expression coincides with the expression given for NMR experiments~\cite{Budker:2013hfa}. 
At the resonance, $\omega_a =  \Omega_{\mathrm{MDM}}$, equation~\ref{eq:sy2a} reduces to
\begin{equation}\label{eq:sy_res}
  S_v(t) =  \frac{\eta_1 \Omega_{\mathrm{EDM}}}{2} \, \cos(\varphi_a) \, t.
\end{equation}
In this case the build-up is linear in time to first order in $\eta_1$.

\begin{figure}
\includegraphics[width=0.9\textwidth]{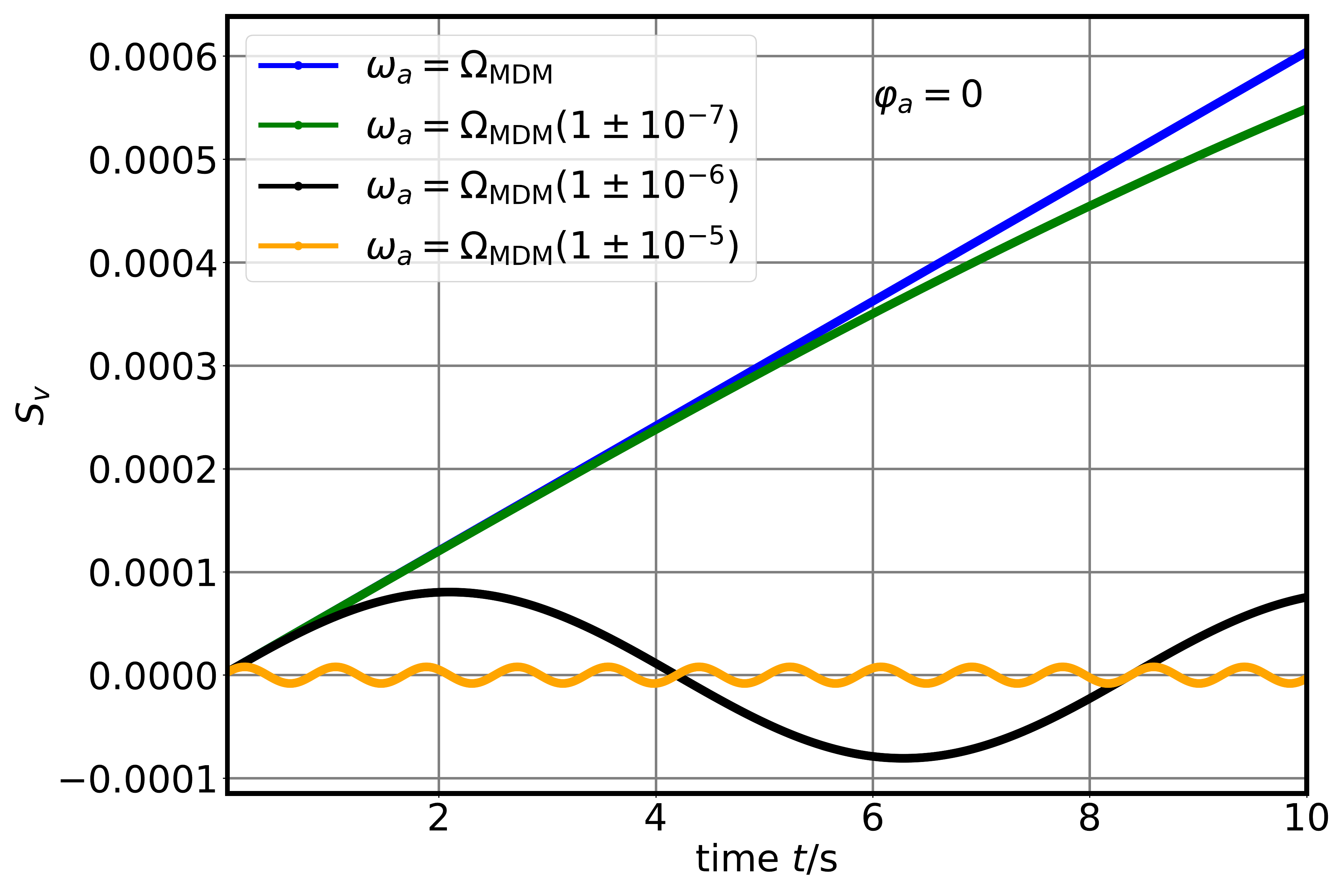}\\
\includegraphics[width=0.9\textwidth]{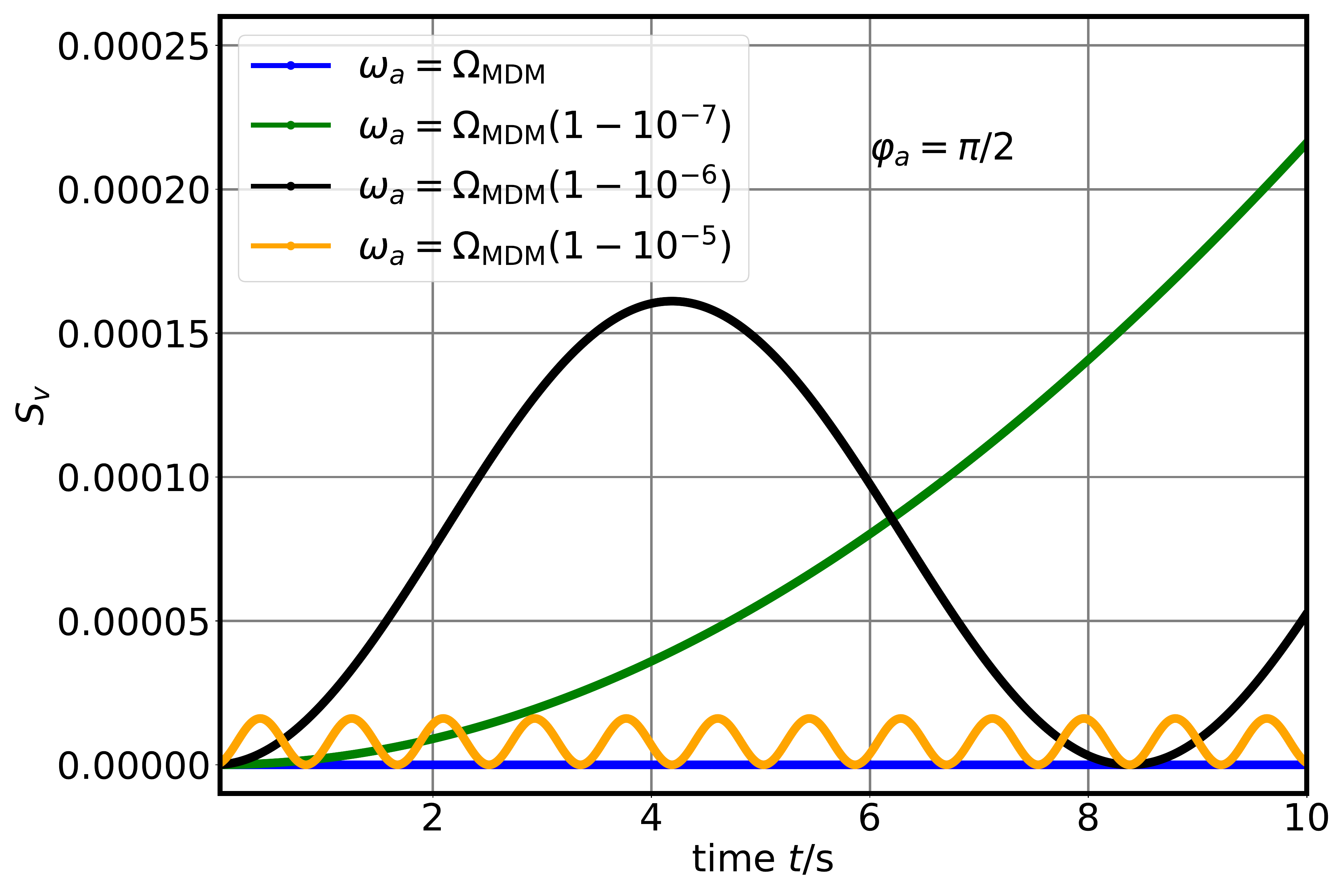}\\
\caption{Vertical spin component $S_v$ as a function of time $t$ 
for $\varphi_a=0$ (upper plot)
and $\varphi_a=\pi/2$ (lower plot) and	
	for different axion 
frequencies $\omega_a$ and
$\Omega_{\mathrm{MDM}}  = 750000.0 \, \si{s^{-1}}$,
$\Omega_{\mathrm{EDM}} \approx 1200000 \, \si{s^{-1}}$,
$\eta_0=0$, $\eta_1=10^{-10}$. \label{fig:sy}}
\end{figure}

%
%\begin{equation}
%S_v(t)=\eta_1 \omega_{\mathrm{EDM}} \left(
%-\frac{ \omega_a}{ (\omega_a - \Omega_{\mathrm{MDM}}) %(\omega_a + \Omega_{\mathrm{MDM}})}  + \frac{ \cos((\omega_a - %\Omega_{\mathrm{MDM}}) t)} {2 (\omega_a - %\Omega_{\mathrm{MDM}}) }                     \right)
%\end{equation}
%
The phase $\varphi_a$ of the axion field is unknown. The experiment should be performed with two bunches in the ring where the polarisations are
orthogonal to each other, which corresponds to two phases $\varphi_a$ and $\varphi_a+\pi/2$. This assures not to miss an axion signal. This can also be seen in  Fig.~\ref{fig:sy}. It shows the build-up of the vertical spin component $S_v$ as a function of time $t$ for $\varphi_a=0$
and $\varphi_a=\pi/2$ and
for different axion frequencies $\omega_a$ and
$\Omega_{\mathrm{MDM}} = 750000.0 \, \si{s^{-1}}$. This $\Omega_{\mathrm{MDM}}$ corresponds to typical 
running conditions
with deuterons of $p=970\,\si{MeV}/c$ at the COoler SYnchrotron COSY of Forschungszentrum J\"ulich
in Germany.
Note that for a given $\varphi_a$ the initial slope is the same independent of $\omega_a$.
One clearly observes the resonance behavior. If $\Omega_{\mathrm{MDM}} = \omega_a$ the polarisation build-up
is maximal for $\varphi_a=0$. The more $\Omega_{\mathrm{MDM}}$ deviates from $\omega_a$, the weaker the signal becomes.

For the special case $\omega_a=0$ equation~\ref{eq:sy} becomes
\begin{equation}
  S_v =  \frac{\Omega_{\mathrm{EDM}}}{\Omega_{\mathrm{MDM}}}   \sin(\Omega_{\mathrm{MDM}} t) \, 
  \left(\eta_0 + \eta_1 \cos(\varphi_a) \right) \, .
\end{equation}
 Compared to equations \ref{eq:sy2} and \ref{eq:sy_res} the signal
 is two times larger. For the following estimates of statistical uncertainties, we continue to use equations \ref{eq:sy2} and \ref{eq:sy_res} for conservative results.

\section{Statistical Error Estimates}\label{sec:staterr}
Equations~\ref{eq:sy2a} and \ref{eq:sy_res} can now be used to calculate statistical sensitivities under various experimental conditions.
We are interested in the error on $\eta_1$.

\subsection{Resonance case}
The best sensitivity is of course given on resonance, i.e. 
$\Omega_{\mathrm{MDM}} = \omega_a$. In this case the spin build-up follows equation~\ref{eq:sy_res}:
\begin{equation}\label{eq:sy3}
 S_v(t) = \eta_1 \frac{\Omega_{\mathrm{EDM}}}{2} \cos(\varphi_a) t \, .
\end{equation}

Assuming that one extracts a beam of $N$ particles continuously on a target with the same rate over a time period $T$
during which the beam polarisation $P$ is assumed to be constant and
using a polarimeter with an average analyzing power $A$ of the scattering process and a fraction $f$ of the beam particles detected, 
the observed vertical polarization (assuming $P_v \ll P$)
will be:
\begin{equation}\label{eq:Pv}
   P_v(t) = P A S_v(t) = P A \eta_1 \frac{\Omega_{\mathrm{EDM}}}{2} \cos(\varphi_a) t \, .
\end{equation}
From this polarization measurement $\eta_1$ can be determined with variance
\begin{equation}\label{eq:var_eta_res}
V(\eta_1) = \left( \frac{1}{\Omega_{\mathrm{EDM}}} \right)^2 \, \frac{96}{fN (ATP \cos(\varphi_a))^2} \, . 
\end{equation}
Details are given in Appendix~\ref{app:var1}.

Adding the information from the two bunches with $\Delta \varphi_a = \pi/2$
one arrives at 
\begin{equation}\label{eq:var_res}
V(\eta_1) = \left( \frac{1}{\Omega_{\mathrm{EDM}}} \right)^2 \, \frac{96}{fN (ATP)^2} \, . 
\end{equation}

\subsection{Off-resonance case}

For the off-resonance case the vertical polarisation is obtained by multiplying equation~\ref{eq:sy2a} with $PA$: 
\begin{equation}~\label{eq:Py}
  P_v(t) =  \eta_1 P A \frac{\Omega_{\mathrm{EDM}}}{2\Delta \omega}
  \left( -\sin(\varphi_a) + \sin(\Delta \omega t + \varphi_a)\right)   \, .
\end{equation}

In order to determine $\eta_1$, the data have to be fitted to the functional form of equation~\ref{eq:Py}.
The three fit parameter are $\eta_1$, $\Delta \omega$ and $\varphi_a$.

\begin{figure}
	\includegraphics[width=\textwidth]{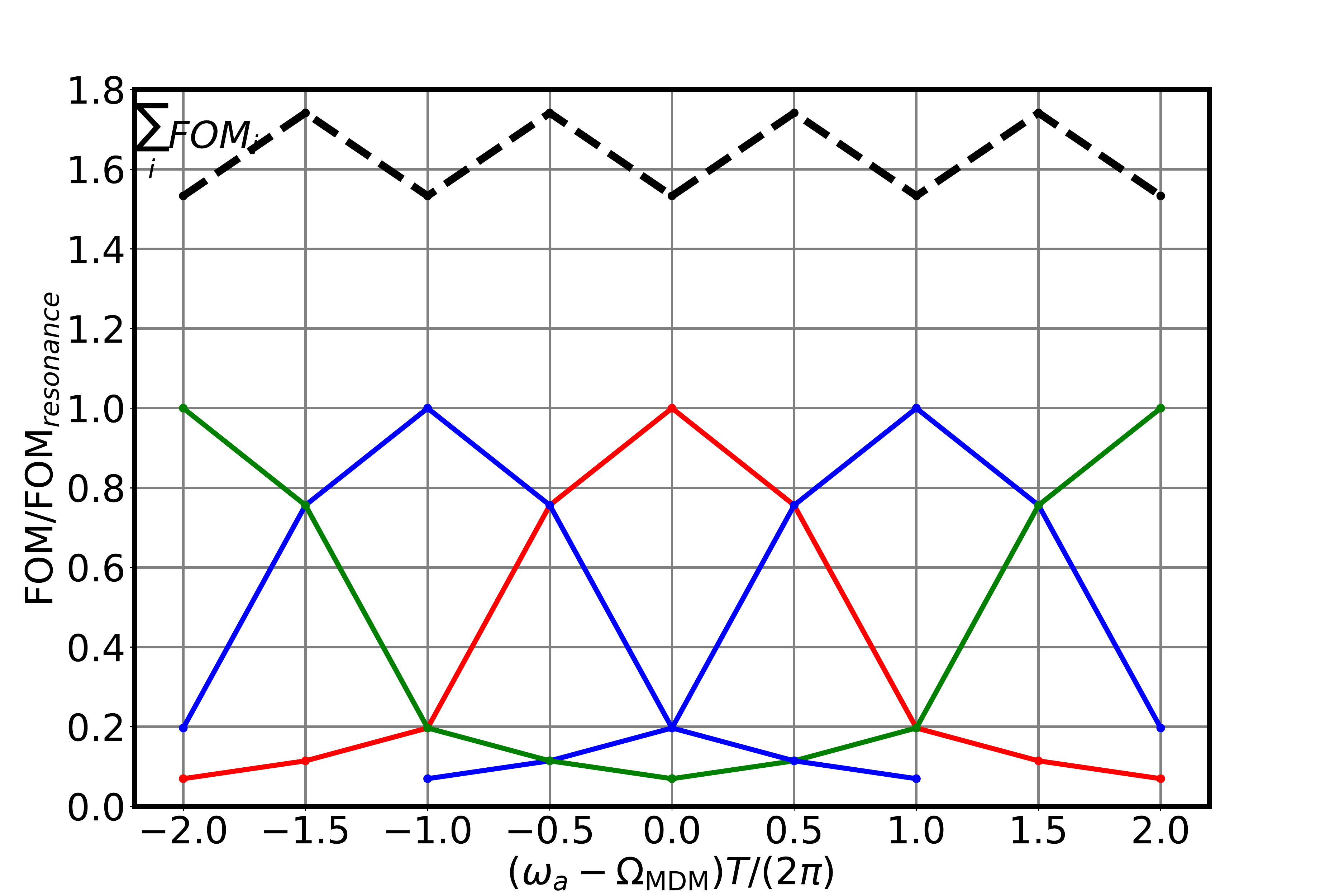}
	\caption{Figure of merit (FOM) as a function of $(\omega_a  - \Omega_{\mathrm{MDM}})T/(2\pi)$ normalized to the FOM at resonance $\Delta \omega = (\omega_a  - \Omega_{\mathrm{MDM}})=0$. 
    Solid lines: FOM for measurements at $\Delta \omega T/(2\pi) = -2,-1,0,1,2$ respectively. Dashed line: sum of FOMs.
		  \label{fig:fom1}}
\end{figure}

The central red curve in Figure~\ref{fig:fom1} shows the figure of merit (FOM) defined as the inverse of the variance of $\eta_1$ as a function of $\Delta \omega T/(2\pi)$ normalized to the FOM at resonance $\Delta \omega= \omega_a  - \Omega_{\mathrm{MDM}}=0$ given by the inverse of equation~\ref{eq:var_res}. If the frequency
is off be $1/T$, with $T$ being the measurement duration, the FOM drops to
roughly 20\%. Details are given in appendix~\ref{app:var2}. This suggests to take measurements separated by $1/T$
in frequency, as indicated by the additional blue and green FOM curves
in Figure~\ref{fig:fom1}.
The upper dashed black curve which is roughly constant shows the sum of the FOMs from the measurements at the different frequencies. 
Experimentally one would not run at frequencies $\Delta \omega T/(2\pi) = \dots,-2 ,-1, 0, 1, 2, \dots$ as indicated in Figure~\ref{fig:fom1} but rather sweep the frequency with the speed (=frequency per time) $v=1/T^2$.

To scan a region of $\Delta f = 1\,\si{kHz}$ with a measurement duration of
 $T=10\,\si{s}$ for a single frequency,
one would thus need a total measurement time
\[ 
\Delta f T^2 = 10^{5} \, \si{s} \, .
\]
In this frequency range $\eta_1$ would be determined with the same accuracy over the whole frequency range.

\subsection{Estimates for the error on the axion-gluon coupling $\frac{C_G}{f_a}$}

According to reference~\cite{Dragos:2019oxn} the relation 
between the EDM $d$ and $\theta_{QCD}$ is given by
$d \approx 10^{-16} \theta_{QCD} e\,\si{cm}$. To simplify the discussion we make no distinction between proton and deuteron.
$\theta_{QCD}$ is connected to the axion field amplitude $a_0$ 
and the axion-gluon coupling strength $C_g/f_a$ via
$\theta_{QCD} = a_0 \, C_g/f_a$. Using the relation between
the axion density $\rho_a$ to the amplitude $a_0 = \sqrt{2 \rho_a}/m_a$
and finally equating 
$\rho_a$ with the local dark matter density
$\rho_{LDM} \approx 0.4 \si{GeV/cm^3} \approx 3 \cdot 10^{-42} \si{GeV^4}$ (see reference \cite{Tanabashi:2018oca}),
assuming that axions saturate the local DM energy,
accuracy estimates for $C_g/f_a$ can be obtained
as a function of the axion mass $m_a$:
\begin{eqnarray}
  d^{osc.} &=& 10^{-16} \, \theta_{QCD} \, e\,\si{cm} \\
    &=& 10^{-16} \, a_0 \, \frac{C_G}{f_a}\\
    &=&  10^{-16} \, \frac{\sqrt{2 \rho_{LDM}}}{m_a} \, \frac{C_G}{f_a}\\
    &=&  2.5 \cdot 10^{-18} \, \frac{C_G}{f_a} \frac{1}{m_a}\,  e\si{V} \, \si{GeV} \, e \si{cm}   = \eta_1 \frac{q\hbar}{2mc} \,S \, .
\end{eqnarray}

Table~\ref{tab:ranges} gives an overview over frequency ranges
accessible at the existing Cooler Synchrotron COSY at Forschungszentrum J\"ulich in Germany
using polarized protons and deuterons
and for a planned prototype storage ring with combined electric and magnetic bending fields
for an EDM measurement~\cite{Abusaif:2018oly}.
Other parameters, like number of stored particles $N$, efficiency $f$, analyzing power $A$, polarization $P$
and spin coherence time $\tau$ are given as well.

%\begin{sidewaystable}[hp!]
	\begin{table}
	%\begin{center}
	\centering
		\begin{tabular}{|l|r|r|r|r|r|r|r|}
			\hline
			&   & \multicolumn{4}{|c|}{COSY} & \multicolumn{2}{|c|}{prototype ring}\\
			\hline
			&  & \multicolumn{2}{|c|}{ proton}           & \multicolumn{2}{|c|}{deuteron}  & \multicolumn{2}{|c|}{proton} \\
			\hline
			momentum & $p/\si{GeV}/c$                  &     0.3      &  3.7    &    0.3    &    3.7     & 0.25      &  0.30    \\
			spin revolution frequency & $\Omega_\mathrm{MDM}$/ $\si{10^6 \, s^{-1}}$ &      5.86     &  72.3    &   0.233   & 2.88     & 7.35 &  0.0 \\
			axion mass & $m_a$/$\si{eV}$  & $4 \cdot 10^{-9}$ & $5\cdot 10^{-8}$ &
			$1.5 \cdot 10^{-10}$ & $2 \cdot 10^{-9}$ &    $5\cdot 10^{-9}$ & 0 \\ 
			magnetic field &$B/\si{T}$          &     0.07  &          0.8      &   0.07  & 0.8       &     0.0  &    0.033            \\
			electric field &$E/\si{MV/m}$    &    $-$    &  $-$     & $-$ & $-$           &       7.4  &    7.4    \\   
			stored particles per bunch & $N$              &  \multicolumn{4}{c|}{$10^{9}$}  &  \multicolumn{2}{c|}{$10^{10}$}  \\
			fraction detected events & $f$              &  \multicolumn{4}{c|}{0.005}   &  \multicolumn{2}{c|}{0.005}  \\
			average analyzing power& $A$              &   \multicolumn{4}{c|}{0.6}   &   \multicolumn{2}{c|}{0.5} \\
			beam polarization & $P$              &   \multicolumn{4}{c|}{0.8}   &   \multicolumn{2}{c|}{0.8} \\
			spin coherence time  &$\tau \,/\si{s}$  & \multicolumn{4}{c|}{1000}    & \multicolumn{2}{c|}{1000}  \\
			\hline
		\end{tabular}
%	\end{center}
	\caption{Parameters used for the estimates. The ring radius of the prototype ring is $R=8.9\,\si{m}$.
		\label{tab:ranges}}
%\end{sidewaystable}
\end{table}
The accuracy estimates are given for two scenarios
\begin{enumerate}
\item One year of beam time ($10^7 \si{s}$) is spent at a single frequency.
\item In one year of beam time a certain range in frequency is covered.
\end{enumerate}

For the duration of a single measurement, we assure that it does not exceed  the axion coherence time, $\tau_{ax}$, given by
\[
     \tau_{ax}  =\frac{\pi \hbar}{m_a} Q
\]
with a quality factor $Q=3 \cdot 10^6$ as in reference~\cite{Chang:2019poy}.

\begin{figure}
\centerline{\includegraphics[width=0.95\textwidth]{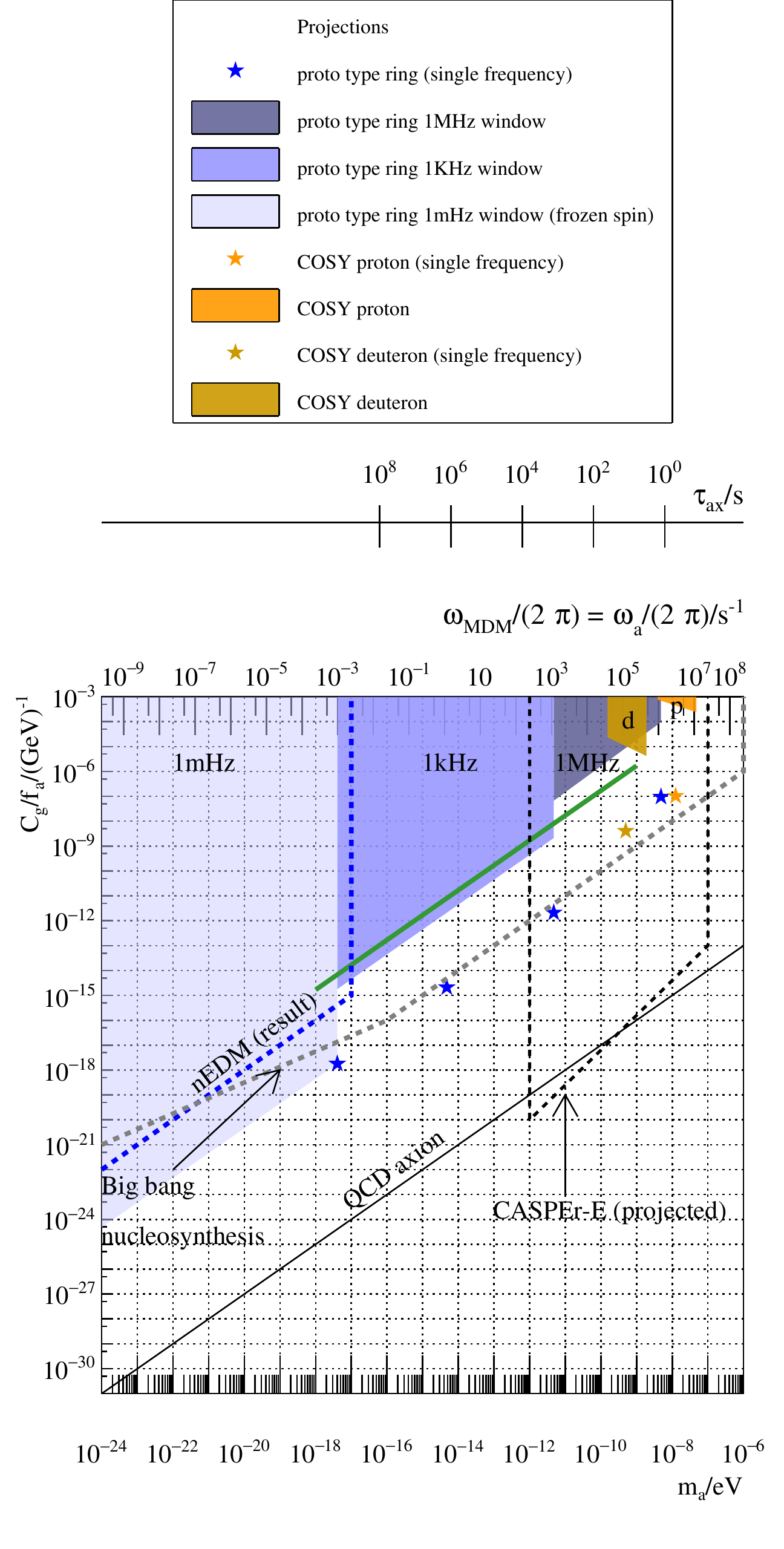}}
\caption{
One $\sigma$ limits for the axion-gluon coupling $C_g/f_a$ reachable within one year running at a fixed frequency (stars)
or over a given frequency range (areas) for COSY (orange) or the prototype ring (blue).
In addition limits reached by the nEDM experiments~\cite{Abel:2017rtm}, nucleosynthesis~\cite{Blum:2014vsa} and prospects for NMR experiments~\cite{Budker:2013hfa} are shown schematically.
The green line shows the estimates obtained in~\cite{Chang:2019poy} with simulations.
\label{fig:limits}}
\end{figure}

The dots in Fig.~\ref{fig:limits} indicate one-$\sigma$
limits one could reach at COSY running with protons or deuterons and for the prototype ring
running at one fixed frequency for one year for each point.

In the second scenario we start with the total running time available in one year, $T_{y} = 10^7\,\si{s}$.
For the prototype ring, if one wants to span a region of $\Delta f=1 \, \si{MHz}$ in one year, 
the duration $T$ is given by 
\[
  T = \sqrt{   \frac{T_y}{\Delta f}    } = 3.2 \,\si{s} \, .
\]
for each frequency interval $\Delta f_i = {1}/{T}$.
%Since the axion coherence time is only about $1\,\si{s}$ at $\omega_a=2\pi \cdot 1\,\si{MHz}$ a length $T=1\,\si{s}$ was chosen.
For a $1\,\si{kHz}$ region, one finds  $T=100 \,\si{s}$.

The corresponding limits are shown in  Fig.~\ref{fig:limits} as colored areas.
The green line shows estimates from reference~\cite{Chang:2019poy} scaled to match them
with the assumptions made in this document about the parameters $N,f,P,A$.

The same is shown for running at COSY.
The fact that the limits using a pure magnetic ring are getting worse at smaller frequency is due to
the fact that
for lower frequencies, the magnetic field is lower, which in turns makes $\Omega_{\mathrm{EDM}}$ 
smaller and
one looses sensitivity. For the combined ring the electric field is constant, a small magnetic field 
is added
to slow down the spin precession. $\Omega_{\mathrm{EDM}}$ varies only very little.

\section{Summary and conclusion}
Analytic expressions for the spin motion in presence of an oscillating EDM in storage rings were 
derived from the Thomas-BMT equation.
These were used to give sensitivity estimates for the axion-gluon coupling at COSY and at a 
prototype EDM ring.
This was done for two scenarios: 1.) Running at one fixed frequency, 2.) covering a wide range in frequency.

The results are in good agreement compared to reference~\cite{Chang:2019poy} where a numerical approach was used to 
find sensitivities.

\section*{Acknowledgments}
This work was supported by the ERC Advanced Grant (srEDM \#694340)
of the European Union and by IBS-R017-D1 of the Republic of Korea.
%Frequency scan, straigh sections
%\begin{acknowledgements}
%\section*{Acknowledgements}
%The authors would like to thank M.~Hartmann for comments and discussions on the paper.
%This work was triggered by discussions on polarimetry for a storage ring electric dipol%e moment (EDM) measurement pursued by the JEDI\footnote{
%\href{http://collaborations.fz-juelich.de/ikp/jedi/}{http://collaborations.fz-juelich.de/ikp/jedi/}} collaboration and was supported by the ERC Advanced Grant (srEDM \#694340)
%of the European Union.
%\end{acknowledgements}

\appendix
 \section{Solution of equation~\ref{eq:dgl}}\label{app:soldgl}
Equation~\ref{eq:dgl} can be written as
\begin{equation}\label{eq:app_dgl}
\dot{\vec S} = (A_{\mathrm{MDM}} + \eta \tilde{A}_{\mathrm{EDM}}(t))  \vec{S} \,. 
\end{equation}
%We note that if the the EDM is time dependent the matrix $ \tilde{A}_{\mathrm{EDM}}(t)$ also depends on time.

To solve equation~\ref{eq:app_dgl} we expand the solution in orders of $\eta$
\begin{equation}\label{eq:S01}
 \vec{S}(t) = \vec{S}_0(t) + \eta \vec{S}_1(t)
\end{equation}

Entering equation~\ref{eq:S01} in equation~\ref{eq:app_dgl} and keeping only terms up to order one in $\eta$ yields
\begin{equation}
  \dot{\vec S}_0 + \eta \dot{\vec S}_1 = A_{\mathrm{MDM}} \vec{S}_0 + \eta (A_{\mathrm{MDM}} \vec{S}_1 + \tilde{A}_{\mathrm{EDM}} \vec{S}_0) \, .
\end{equation}
Thus
\begin{eqnarray}
  \dot{\vec S}_0 &=& A_{\mathrm{MDM}} \vec{S}_0   \, , \label{eq:S0}\\
  \dot{\vec S}_1 &=& (A_{\mathrm{MDM}} \vec{S}_1 + \tilde{A}_{\mathrm{EDM}} \vec{S}_0)  \label{eq:eta1} \, .
\end{eqnarray}

Since $A_{\mathrm{MDM}}$ does not depend on $t$, equation~\ref{eq:S0} has the solution
\begin{equation}
  \vec{S}_0(t) = \mathrm{exp}(A_{\mathrm{MDM}} t) \vec{S}(0)
\end{equation}
with arbitrary initial condition $\vec{S}(0)$.

The solution for the equation~\ref{eq:eta1} can be found using the variation of constant method:
\begin{equation}
  S_1 = \mathrm{exp}(A_{\mathrm{MDM}} t) S(0) +  \int_0^t  \mathrm{exp}(A_{\mathrm{MDM}}(t-s)) \tilde{A}_{\mathrm{EDM}} S_0(t)\dd s \, .
\end{equation}% Duhamel's formula

Up to first order in $\eta$ the solution is
\begin{eqnarray}
  \vec{S}(t) &=& \vec{S}_0(t) + \eta \vec{S}_1(t) \\
    &=& (1+\eta) \, \mathrm{exp}(A_{\mathrm{MDM}} t) \, \vec{S}(0) + \eta \, \int_0^t  \mathrm{exp}(t-s) \tilde{A}_{\mathrm{EDM}}  \mathrm{exp}(A_{\mathrm{MDM}} t) \, \vec{S}(0) \dd s
\end{eqnarray}

Using Mathematica~\cite{Mathematica} one finds $\vec{S}(t) = A(t)  \vec{S}(0)$ with
\footnotesize
\begin{eqnarray}
A_{11} &=& (1+\eta_0) \cos(\Omega_{\mathrm{MDM}} t) \hspace{8cm}\\
A_{12} &=& \frac{\eta_0 \Omega_{\mathrm{EDM}} (\cos(\Omega_{\mathrm{MDM}} t)-1)}{\Omega_{\mathrm{MDM}}} \nonumber  \\
   &+&\,\eta_1 \Omega_{\mathrm{EDM}} \,\Big( \frac{
   (\sin(\varphi_a) (\omega_a \sin (\Omega_{\mathrm{MDM}}t)-
   \Omega_{\mathrm{MDM}} \sin (\omega_a t))}{\omega_a^2-\Omega_{\mathrm{MDM}}^2} \nonumber \\
  && \hspace{1.5cm} +\frac{\Omega_{\mathrm{MDM}} \cos (\varphi_a) (\cos (\omega_a t)-\cos(\Omega_{\mathrm{MDM}} t))}{\omega_a^2-\Omega_{\mathrm{MDM}}^2} \Big)\\
 A_{13} &=&  (1+\eta_0) \sin (\Omega_{\mathrm{MDM}} t) \\
A_{21} &=& \frac{\eta_0 \Omega_{\mathrm{EDM}} (\cos (\Omega_{\mathrm{MDM}}
   t)-1)}{\Omega_{\mathrm{MDM}}} \nonumber  \\
    &-& \eta_1 \Omega_{\mathrm{EDM}} \Big( \frac{
   ( \cos ((\omega_a
   -\Omega_{\mathrm{MDM}})
   t+\varphi_a)}{2 (\omega_a-\Omega_{\mathrm{MDM}})
   } \nonumber \\
   && \hspace{1.5cm} +\frac{(\Omega_{\mathrm{MDM}}-\omega_a) \cos (
   (\omega_a+\Omega_{\mathrm{MDM}})t+\varphi_a)-2 \Omega_{\mathrm{MDM}}
   \cos (\varphi_a))}{2 (\omega_a-\Omega_{\mathrm{MDM}})
  (\omega_a+\Omega_{\mathrm{MDM}})} \Big)\\
 A_{22} &=& 1+\eta_0 \\
 A_{23} &=&  \frac{\eta_0 \Omega_{\mathrm{EDM}} \sin (\Omega_{\mathrm{MDM}}
   t)}{\Omega_{\mathrm{MDM}}} \nonumber \\
    &+& \eta_1 \Omega_{\mathrm{EDM}} \Big( \frac{
   ( \sin ((\omega_a
   -\Omega_{\mathrm{MDM}})
   t+\varphi_a)}{2 (\omega_a-\Omega_{\mathrm{MDM}})
   } \nonumber \\
   && \hspace{1.5cm} + \frac{(\omega_a-\Omega_{\mathrm{MDM}}) \sin (
   (\omega_a+\Omega_{\mathrm{MDM}})t+\varphi_a)-2 \omega_a
   \sin (\varphi_a))}{2 (\omega_a-\Omega_{\mathrm{MDM}})
   (\omega_a+\Omega_{\mathrm{MDM}})} \Big) \label{eq:A23}\\
 A_{31} &=& -(1+\eta_0) \sin (\Omega_{\mathrm{MDM}} t) \\  
A_{32} &=&  -\frac{\eta_0
	\Omega_{\mathrm{EDM}} \sin (\Omega_{\mathrm{MDM}} t)}{\Omega_{\mathrm{MDM}}} \nonumber \\
   &+&\eta_1 \Omega_{\mathrm{EDM}} \, \Big( \frac{ (\omega_a \sin
   (\varphi_a) \cos (\Omega_{\mathrm{MDM}} t)-\omega_a \sin
   (\omega_a t+\varphi_a)}{(\omega_a-\Omega_{\mathrm{MDM}})
   (\omega_a +\Omega_{\mathrm{MDM}})} \nonumber \\
  && \hspace{1.5cm} +\frac{\Omega_{\mathrm{MDM}} \cos
   (\varphi_a) \sin (\Omega_{\mathrm{MDM}}
   t))}{(\omega_a-\Omega_{\mathrm{MDM}})
   (\omega_a +\Omega_{\mathrm{MDM}})} \Big)\\
A_{33} &=& (1+\eta_0) \cos (\Omega_{\mathrm{MDM}} t) 
\end{eqnarray}
\normalsize
Note that $\eta = \eta_0 + \eta_1 \cos(\omega_a t + \varphi_a)$.
We are mainly interested in the entries $A_{23}$ and $A_{21}$ which gives the vertical polarisation
in case of an initial in plane polarisation.

 \section{Variance on $\eta_1$}\label{app:var}

\subsection{Resonance case: variance of a slope}\label{app:var1}
Starting point is equation~\ref{eq:Pv}
\begin{equation}\label{eq:Pv1}
   P_v(t) = P A S_v(t) = P A \eta_1 \frac{\Omega_{\mathrm{EDM}}}{2}  \cos(\varphi_a)\, t.
\end{equation}

The variance on the slope parameter $s = P A \eta_1 \frac{\Omega_{\mathrm{EDM}}}{2} \cos(\varphi_a)$
of a straight line is
\[
  V(s) = \frac{\sigma^2}{N_{\mathrm{points}} V(t)} \, ,
\]
where $\sigma$ is the error on each individual point where the curve is measured.
$N_{points}$ is the number of points entering the fit and $V(t)$ is 
the variance of the points along the time axis.
For evenly distributed values in a time interval $T$, one has $V(t) = T^2/12$.
If the polarization is determined from an azimuthal asymmetry
one has~\cite{Pretz:2018bze}:
\[
  \sigma^2 = \frac{2}{n} \, ,
\]
where $n$ is the number of events entering the analysis for a single point.
Evidently for the total number of events one has
$Nf = n N_{points}$.

Putting everything together one finds 
\begin{equation}
  V(s) = \frac{24}{fN T^2} \, .
\end{equation}

Translated to the variance on $\eta_1$ one finds the expression given in equation~\ref{eq:var_res}
\begin{equation}\label{eq:var_eta1}
  V(\eta_1) = \frac{24}{fN (PAT \cos(\varphi_a))^2} \left( \frac{2}{\Omega_{\mathrm{EDM}}} \right)^2 \, .
\end{equation}

\subsection{Off-resonance case: variance of an amplitude}\label{app:var2}

%We consider a counting rate of the form
A polarization given by equation~\ref{eq:Py} leads to the following count rate in the detector:
\begin{equation}\label{eq:N}
  N(t) \propto 1 + \eta_1 \frac{PA \Omega_{\mathrm{EDM}}}{2 \Delta \omega} \, \left(-\sin(\varphi_a) + \sin(\Delta \omega t +\varphi_a) \right) \cos(\Phi)
\end{equation}
where $\Phi$ is the azimuthal angle of the scattered particle.
%The factor $B$ is given by
%\begin{equation}
%  B = PA \eta_1 \frac{\Omega_{\mathrm{EDM}}}{2} \, .
%\end{equation}
There are three unknowns $\eta_1$, $\Delta \omega$ and $\varphi_a$.
To estimate the uncertainty on $\eta_1$ we consider the extended maximum likelihood method applied
to the counting rate in equation~\ref{eq:N}.
The log-likelihood function $\ell$ has the form
\begin{equation}
  \ell = \sum_{i=1}^{N_{\mathrm{events}}} \log\left(1 + \eta_1 \frac{PA \Omega_{\mathrm{EDM}}}{2 \Delta \omega} \left(-\sin(\varphi_a) + \sin(\Delta \omega t +\varphi_A) \right) \cos(\Phi_i) \right)- \log(N_{tot})\, ,
\end{equation}
where $N_{tot}$ is the total number of events detected.

To get the covariance matrix for the three unknowns $\eta_1, \Delta \omega$ and $\varphi_a$ one has to consider the expectation values of the second derivatives of the likelihood function.

The the second derivative with respect to $\eta_1$ it is for example given
by
\begin{equation}
	\avrg{\frac{\partial^2 \ell}{\partial \eta_1^2}}
	= \int_0^T \frac{\partial^2 \ell}{\partial \eta_1^2} N(t) \mathrm{d}t \, .
\end{equation}

For $\eta_1 /\Delta \omega \ll 1$ and a measurement time $T=2\pi/\Delta \omega$ (corresponding roughly to the black curves in Figure~\ref{fig:sy}), one finds for example for the error on $\eta_1$:
\begin{eqnarray}
\sigma_{\varphi_a=0}^2 &=& \frac{1}{(\Omega_{\mathrm{EDM}} ATP)^2 fN} \, \frac{128 \pi^2 (15 + 2 \pi^2)}{(3 + 4 \pi^2)} \approx \frac{1033}{(\Omega_{\mathrm{EDM}} ATP)^2 fN} \\
 & & \quad \mbox{for} \, \, \varphi_a = 0 \nonumber \, , \\
\sigma_{\varphi_a=\pi/2}^2 &=& \frac{1}{(\Omega_{\mathrm{EDM}} ATP)^2 fN} \,\frac{128 \pi^2 (15 - 2 \pi^2)}{33 - 4 \pi^2} \approx \frac{924}{(\Omega_{\mathrm{EDM}} ATP)^2 fN} \,\\ 
  & & \quad \mbox{for} \, \, \varphi_a = \frac{\pi}{2} \, .\nonumber
\end{eqnarray}
Combining these two measurements leads to a
\begin{equation}
V(\eta_1) =  \frac{488}{(\Omega_{\mathrm{EDM}} ATP)^2 fN} \
\end{equation}
which is approximately a factor 5 larger compared to the resonance case 
in equation~\ref{eq:var_eta1}.
%In case the vertical polarisation varies according to 
%\[
%   B \sin(\omega t) 
%\]
%the oberserved counting rate in the detector is
%\[
%   n(t,\varphi) \propto 1 + B \sin(\omega t) \sin(\varphi)
%\]
%where $\varphi$ denotes the azimuthal angle.

%The log-likelihood function is given by
%\[
%  \ell = \ln(\mathcal{L}) = \sum_{i=1}^M \ln(1 + B \sin(\omega t)  \sin(\varphi))
%\]
%$M=fN$ denotes the number of detected events.
%(Remember that $N$ is the number of beam particles and $f$ is the fraction of detected particles.)

%For the expectation value of the second derivative one finds for $B \ll 1$
%\[
%   \avrg{ \frac{\partial^2 \ell}{\partial B^2}} \approx - \frac{M}{2 \pi T}  \int_0^{2\pi} \int_0^T \sin^2(\omega t) \sin^2(\varphi) d%t d\varphi
%   \approx  -\frac{M}{4} \, ,
%\]
%if $T \gtrsim  2\pi/\omega$. 
%The variance of $B$ is then given
%by
% \[
%V(B) = \frac{4}{Nf} \, .
%\]

%Translating this to the variance of $\eta_1$ leads to
%\begin{equation}
%V(\eta_1) = \frac{4}{fN} \frac{4 (\omega_a  - \Omega_{\mathrm{MDM}})^2}{(PA \Omega_{\mathrm{EDM}})^2} \, .
%\end{equation}

\bibliography{literature_axion.bib,literature_edm.bib}

\bibliographystyle{ieeetr}

\end{document}